\documentclass[10pt,journal,compsoc]{IEEEtran}
\ifCLASSOPTIONcompsoc
  \usepackage[nocompress]{cite}
\else
  \usepackage{cite}
\fi
\ifCLASSINFOpdf
   \usepackage[pdftex]{graphicx}
   \graphicspath{{images/}}
   \DeclareGraphicsExtensions{.pdf,.jpeg,.png}
\else
\fi
\usepackage{amsmath}
\ifCLASSOPTIONcompsoc
 \usepackage[caption=false,font=footnotesize,labelfont=sf,textfont=sf]{subfig}
\else
 \usepackage[caption=false,font=footnotesize]{subfig}
\fi

\usepackage{stfloats}
\usepackage[normalem]{ulem}
\usepackage{xcolor}
\usepackage{paralist}
\usepackage{scalerel}
\usepackage{hyperref}
\usepackage{bm}
\usepackage[locale=FR]{siunitx}
\usepackage{verbatim}

\usepackage{multicol}
\newif\ifShowEdits

\ShowEditsfalse 

\ifShowEdits
    \definecolor{blue}{rgb}{0,0,1}
    \renewcommand{\sout}[1]{\unskip}
\else
    \definecolor{blue}{rgb}{0,0,0}
    \renewcommand{\sout}[1]{\unskip}
\fi

\begin{document}
%
\title{An Integrated Visual Analytics System for Studying Clinical Carotid Artery Plaques}

%
%
%
%

\author{
    Chaoqing~Xu,
    Zhentao~Zheng, 
    Yiting~Fu, 
    Baofeng~Chang,
    Legao~Chen,
    Minghui~Wu,
    ~Mingli~Song,
    and~Jinsong~Jiang
    
    \IEEEcompsocitemizethanks{\IEEEcompsocthanksitem C. Xu, Z. Zheng, and M. Wu are with Computer and Computing Science, Hangzhou City University,Hangzhou,Zhejiang 310000,China.\protect\\
    E-mail: xucq@hzcu.edu.cn,32101157@stu.hzcu.edu.cn.
    \IEEEcompsocthanksitem Y. Fu is with Zhejiang AI Healthcare Innovation Center, Hangzhou, Zhejiang,China.\protect\\
    E-mail: fyt@zjesshine.cn
    \IEEEcompsocthanksitem B. Chang is with Hangzhou Dianzi University, Hangzhou, Zhejiang, China.\protect\\
    E-mail: baofeng.chang@foxmail.com
    \IEEEcompsocthanksitem L. Chen, J. Jiang are with General Surgery, Cancer Center, Department of Vascular Surgery, Zhejiang Provincial People’s Hospital,HangZhou, ZheJiang, China.\protect\\
    E-mail: chenlegao2022@126.com,jiangjinsong2022@126.com
     \IEEEcompsocthanksitem M.Song is with Computer Science and Technology, Zhejiang University, HangZhou, Zhejiang, China.\protect\\
    E-mail: songml@zju.edu.cn
   }
   
    \thanks{Manuscript received September xx, 2020; revised xxx, 2020.}
}
%

%

\markboth{Journal of \LaTeX\ Class Files,~Vol.~14, No.~8, August~2023}%
{Shell \MakeLowercase{\textit{et al.}}: Bare Demo of IEEEtran.cls for Computer Society Journals}

\IEEEtitleabstractindextext{%
\begin{abstract}
Carotid artery plaques can cause arterial vascular diseases such as stroke and myocardial infarction, posing a severe threat to human life. However, the current clinical examination mainly relies on a direct assessment by physicians of patients' clinical indicators and medical images, lacking an integrated visualization tool for analyzing the influencing factors and composition of carotid artery plaques. We have designed an intelligent carotid artery plaque visual analysis system for vascular surgery experts to comprehensively analyze the clinical physiological and imaging indicators of carotid artery diseases. The system mainly includes two functions: First, it displays the correlation between carotid artery plaque and various factors through a series of information visualization methods and integrates the analysis of patient physiological indicator data. Second, it enhances the interface guidance analysis of the inherent correlation between the components of carotid artery plaque through machine learning and displays the spatial distribution of the plaque on medical images. Additionally, we conducted two case studies on carotid artery plaques using real data obtained from a hospital, and the results indicate that our designed carotid analysis system can effectively provide clinical diagnosis and treatment guidance for vascular surgeons.
\end{abstract}

}

\maketitle

\IEEEdisplaynontitleabstractindextext

%
\IEEEpeerreviewmaketitle

\IEEEraisesectionheading{\section{Introduction}\label{sec:introduction}}

\textcolor{blue}{Atherosclerosis (AS) is an arterial disease caused by various factors, leading to ailments such as stroke and myocardial infarction~\cite{saba2021multimodality}, which pose serious threats to human life. The narrowing of the lumen caused by carotid artery plaques is a significant mechanism leading to cerebral ischemia~\cite{bonati2022management}, making the analysis of carotid artery plaques of great value in clinical research.}

Currently, in clinical practice, doctors usually evaluate patients' physiological indicator data by looking at the text, including demographics, hypertension, diabetes, smoking history, troponin, blood routine, and other factors. However, there is a lack of integrated analysis tools, which greatly limits the efficiency of carotid artery plaque evaluation. In clinical practice, the imaging of carotid artery plaque is mainly done through ultrasound, CTA, and MRA. However, these imaging methods can only show visual information such as the size and degree of stenosis. Although researchers have done a lot of work based on machine learning to improve the accuracy of plaque classification in terms of plaque composition and plaque segmentation~\cite{liu2022deep}
, there is currently no intelligent tool to provide effective information about the pathological mechanisms of carotid artery plaque. Therefore, further research and development of carotid artery plaque visual analysis tools are needed to more accurately evaluate the pathological process and manifestations of carotid artery plaque.

There are many challenges in the analysis of carotid artery plaque at present. First, the formation and development of carotid artery plaque are affected by various factors, such as vascular endothelial cell damage, lipid metabolism disorders, hypertension, diabetes, etc., and there are complex interactions between these factors.
How to efficiently analyze the impact of these factors on carotid artery plaque is an important issue. Second, the prognosis and stroke risks of carotid artery plaque are related to many factors, such as the size of the plaque, the internal morphological characteristics of the plaque, and the location of the plaque. At the same time, there are interactions and uncertainties between these factors, making it more difficult to predict the prognosis of carotid artery plaque. How to effectively analyze the intrinsic correlation of different types of plaques is also an urgent problem to be solved.

\textcolor{blue}{Standalone machine learning can model complex arotid
artery plaque data relationships, however, the lack of interpretability is a significant obstacle. Additionally, iterative, exploratory analysis that may not be achievable solely through machine learning. Visualization is a highly effective means for analyzing multifactorial complex interactions, effectively presenting the risk factors of carotid artery plaque formation and development visually. This allows vascular surgery experts to intuitively understand pathological manifestations and interactively explore potential pathological mechanisms. }

Based on the above challenges and consultation with vascular surgery experts, we then designed an intelligent carotid artery plaque visual analysis system. To our knowledge, this is the first carotid artery plaque visualization system that integrates clinical physiological indicators and radiomic features for vascular surgery experts to diagnose and evaluate efficiently. By integrating multiclass SVM model, the system can intelligently analyze the correlation of different plaque types and evaluate the importance of radiomic features. 
\textcolor{blue}{The system assists experts in directly observing the patterns among patient groups, as well as the distribution of features highly correlated with the disease, by jointly analyzing physiological indicators and radiomic features through visualization charts. } The contributions of this article are as follows:

\begin{compactitem}
	\item A customized visualization system for carotid artery plaques, which can be used by vascular surgeons for visual analysis of clinical data on atherosclerosis.
	\item A workflow that integrates clinical physiological indicators and radiomic features of carotid artery plaques, which can be used to explore the pathogenesis of atherosclerosis.
	\item A real case analysis of carotid artery plaques, demonstrating the system's exploratory and interactive capabilities and providing guidance for the clinical diagnosis and treatment of atherosclerosis.
\end{compactitem}

\section{Related Work}
\label{sec:backgroundAndRelatedWork}

\begin{color}{blue}


    \subsection{Carotid Artery Plaques Research}
In recent years, machine learning methods have been widely applied to carotid Artery plaque research. 
Related tasks mainly focus on plaque segmentation and classification. For instance, Loizou et al. proposed a semi-automatic method that utilizes various snake models 
to segment arterial plaques from 2D Doppler images~\cite{yinyue2011thesis}, while Bonanno et al. employed torFlow  snake models to segment plaques, achieving satisfactory segmentation results 
~\cite{bonanno2017automatic}. Wei et al. utilized deep residual networks (ResNet) to automatically extract features from carotid ultrasound images, and to classify and evaluate the performance of three different regions of interest\cite{ma2019plaque}. In terms of method types, related research is primarily concentrated on automatic or semi-automatic Total Plaque Area (TPA) segmentation and measurement~\cite{loizou2007integrated, loizou2014integrated}. UNet++ based methods have demonstrated superior performance in this field of research~\cite{zhou2021deep}. LucaSaba et al. designed characterizations based on multiple artificial intelligence models for component analysis and vulnerability evaluation of carotid ultrasound plaques. From the data perspective, as B-ultrasound, MRA, and CTA imaging can provide clinical doctors with features such as location, morphology, and size, they are the most widely used data in current clinical research, improving the diagnostic efficiency of carotid Artery plaque for clinical doctors~\cite{fu2023deep}. 

Although the aforementioned research has made significant progress, it merely analyses based on imaging indicators. 
This study explores the inherent connections of multiple factors of carotid  artery plaque through visual interaction means by integrating patient clinical indicators and medical imaging for the fusion analysis of carotid diseases. 

\subsection{Medical Image Visualization}

Medical image visualization is a crucial field aimed at improving and enhancing the comprehension, interpretation, and presentation of medical imaging data. Volume rendering is a common medical image visualization technique that transforms three-dimensional data into two-dimensional images, allowing doctors to observe more intuitive images and better understand and interpret internal structures. 
From the perspective of technological development, related techniques mainly focus on enhancing image rendering effects~\cite{9665344}, optimizing computational efficiency~\cite{9903564}, and developing new interaction methods~\cite{10091196}, thereby achieving real-time high-quality volume rendering~\cite{8440063,7539334}. In addition, the medical imaging volume rendering technique, combined with deep learning, is playing an increasingly significant role. For instance, Wang et al. proposed a novel visualization-guided computation paradigm, combining direct 3D volume processing and volume rendering cues to better capture small/micro-structures~\cite{9222053}. Wang et al. proposed the DeepOrganNet architecture based on deep learning for real-time generation and visualization of fully high-fidelity 3D/4D organ geometrical models from complex medical images, significantly reducing surgical time to achieve real-time visualization~\cite{8809843}. From an application perspective, volume rendering technology can help doctors view and understand the three-dimensional structure of tissues more accurately, thereby making more precise disease diagnoses and treatment plans~\cite{8842614}. 

Despite significant advancements in the field of medical image visualization, there remains substantial potential for further development. This paper does not focus on innovating volume rendering algorithms; instead, it emphasizes application research aimed at enhancing medical image analysis. By integrating two-dimensional images with rendering effects of carotid artery plaque components, it provides vascular surgery experts with an intuitive understanding of the spatial structure of carotid artery plaques.

\end{color}

\section{Design Goals}
\label{sec:goals}

\begin{color}{blue}

The target users of this system are vascular surgery physicians.To ensure that the system aligns with actual clinical needs, we visited the hospital monthly to engage in face-to-face discussions with these experts, delving deeply into the challenges they encounter in their daily work. We not only listened to their concerns but also analyzed the data they had, collaboratively determining the system's design goals to meet real-world requirements and expectations. Through nearly a year of continuous communication and repeated deliberations, we not only clarified the design goals but also adjusted and refined the research plan multiple times. The design goals are as follows:

\vspace{3pt}
\noindent\textbf{DG 1: Display of CTA images and plaques.}

The intuitive display of CTA images and carotid plaques helps experts make an intuitive assessment of the disease. According to the requirements of the experts, the designed interface needs to display the original CTA images and the imaging of plaques on both sides of the carotid artery, and mark them with different colors according to their composition.

\vspace{3pt}
\noindent\textbf{DG 2: Analysis of clinical physiological indicators.}

Vascular surgeons lack a systematic understanding of clinical physiological indicators and cannot fully understand the overall status of patients. Further guiding experts to efficiently analyze clinical physiological indicators can improve diagnostic efficiency and formulate targeted treatment plans.

\vspace{3pt}
\noindent\textbf{DG 3: Analysis of radiomic features.}

Vascular surgeons lack knowledge about the intrinsic correlation and pathological mechanisms of the components of carotid plaques. By analyzing radiomic features of carotid plaques, potential disease patterns can be explored, providing new insights into the formation, development, and transformation of the disease for domain experts.

\vspace{3pt}
\noindent\textbf{DG 4: Visual interactivity of the system.}

Given the large volume of clinical physiological indicators and carotid plaque data, another requirement of vascular surgeons is the visual interactivity of the system. By allowing for measurement and scaling of images, as well as interactive display of data, they can effectively analyze and evaluate the disease.

\end{color}

\section{Methods} 
\label{sec:methods}

\subsection{System Overview}
\label{sec:systemoverview}

\begin{figure*}
    \centering
    \includegraphics[width=\linewidth]{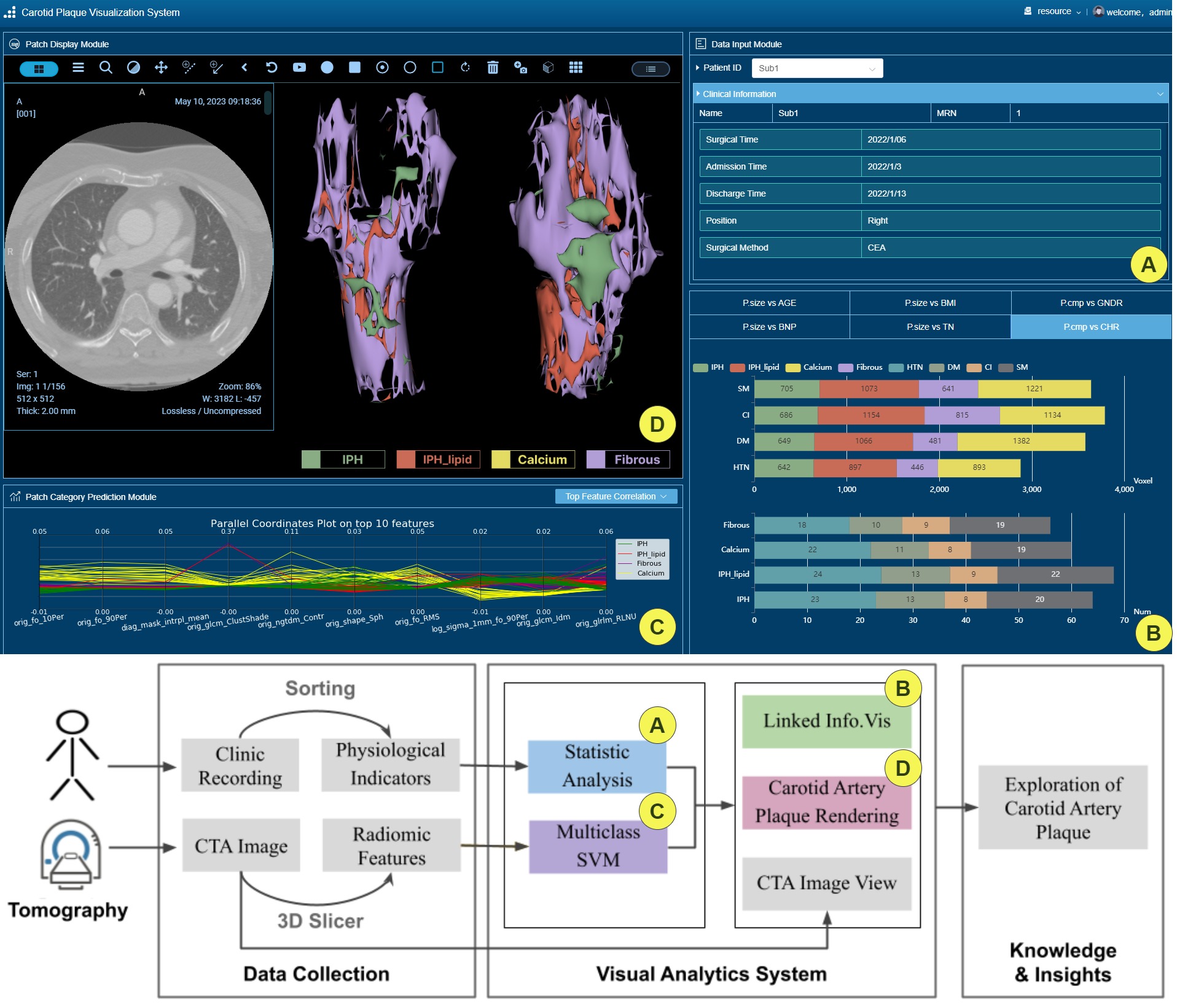}
    \caption{The UI of our VA system (top) and workflow (bottom), where each analysis step is annotated with the label of the corresponding system
module.It includes four main elements: 
(A) Data Input Module.
(B) Linked Information Visualization Module.
(C) Multicalss Classification Module.
(D) Carotid Artery Plaque Rendering Module.
 }
    \label{fig:main}
\end{figure*}

\begin{color}{blue}
Fig.\ref{fig:main} shows our Visual Analytics (VA) workflow and User Interface (UI), which includes the collection of clinical data and CTA images. The use of the system begins by selecting cohorts in Fig.\ref{fig:main}A, and the basic clinical information of the selected sample is displayed in the drop-down box. Next, in Fig.\ref{fig:main}B, multiple views of the cohort data is carried out to reveal the correlations of demographic and various clinical physiological related to carotid artery plaques. Subsequently, by conducting multi-classification training on the radiomic features, we obtain the importance ranking of different plaque components and radiomic features, as seen in Fig.\ref{fig:main}C. It also supports the cooperative analysis of radiomic features and chronic disease. Finally, by selecting samples of interest, the individual's CTA images and spatial distribution of carotid artery plaques are displayed in Fig.\ref{fig:main}D, assisting domain experts in gaining insights into the carotid artery.
\end{color}

\begin{figure}
    \centering
    \includegraphics[width=\linewidth]{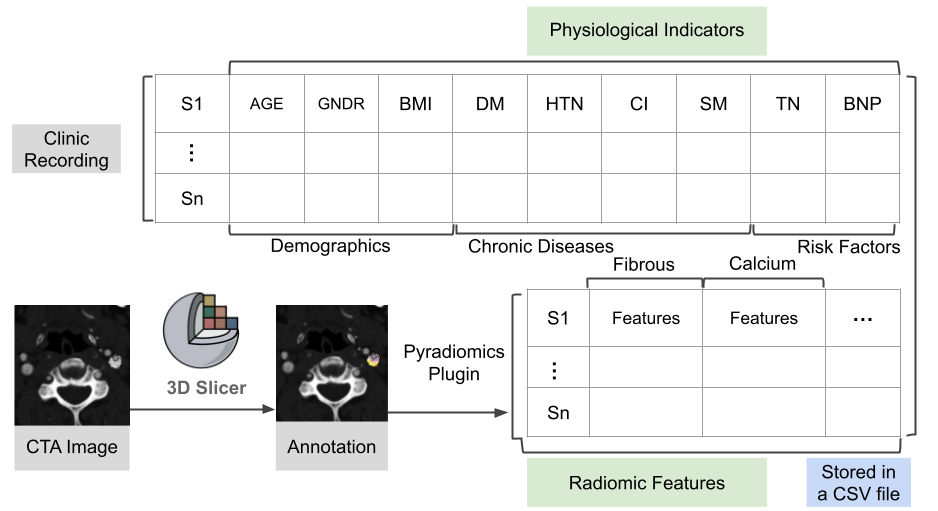}
    \caption{\textcolor{blue}{Radiomic feature extraction and the cohort data stored in a
CSV format.}}
    \label{fig:Data_Description}
\end{figure}

\subsection{Data Description}
\label{sec:dataDescription}
The data for this system is sourced from the patient database at Zhejiang Provincial People's Hospital. It primarily consists of two types of data: clinical data and plaque CTA image data.

\begin{color}{blue}
\vspace{3pt}
\noindent\textbf{1) Clinical Recording Data.}
Clinical recording data primarily include patients' demographics (age, gender, BMI), chronic disease information, such as hypertension (HTN), diabetes mellitus (DM), cerebral infarction (Cl), and smoking history (SM), relevant risk factors, like troponin (TN) and B-type natriuretic peptide (BNP), and a series of symptom information, etc. These are closely related to the incidence and severity of plaques~\cite{wang2017risk}. As biochemical markers of cardiovascular diseases, they reflect the biological and physiological status of the heart~\cite{eggers2008prevalence}. 
Notably, to protect patient privacy, the data have been anonymized, with patient names and medical record numbers concealed. The formation is shown in Fig\ref{fig:Data_Description}(Top).

\vspace{3pt}
\noindent\textbf{2) CTA Imaging Data.}
CTA imaging data mainly refers to the DICOM image cohorts of the patients' head and neck portion after CTA scanning. We first invited experienced vascular surgeons to manually annotate the carotid artery plaque using 3D Slicer, an integrated medical image processing software that provides image registration, annotation, and visualization. According to the manifestations of atherosclerotic lesions and doctors' experience, the carotid artery plaques were annotated into four types: Intraplaque Hemorrhage (IPH), IPH with high lipid content (IPH\_lipid), Calcium tissue, and Fibrous tissue. 
After the annotation, we utilized the "Pyradiomics" plugin of 3D Slicer, an open-source software package for radiomic feature extraction, to extract radiomic features for each type of plaque component. 
Additionally, to carry out potential differences analysis between the left and right carotid arteries of patients, we also differentiated features of both sides. Fig.\ref{fig:Data_Description}(bottom) displays the radiomic feature acquisition process.

\vspace{3pt}
\noindent\textbf{3) Cohort Formation.}
We integrate clinical physiological indicators and radiomic features, form a matrix representing all carotid artery plaques with sample IDs as columns and features as rows, and store it as a CSV file. The formulated cohort data is used in the ML learning pipeline in the system. Note that we only use the label and extracted features to train the ML models, while the demographics and visit dates provide context when displaying the ML results.

\end{color}

\subsection{System Modules}
\label{sec:System module description}

\subsubsection{Data Input Module}
\label{sec:plaque Display Module}
\textcolor{blue}{This module primarily displays the patient's clinic information, and vascular surgeons can select patients of interest through clicking patients' ID from the patient list.  Details of clinic information is described in Sec.\ref{sec:dataDescription}, which are relevant factors in the development of carotid artery plaques. In addition, the clinic information also contains patient's admission and discharge time, surgical time, plaque location, and surgical methods. They are also important references for assessing basic disease conditions, predicting prognosis, and formulating treatment plans. These clinical details help doctors understand the patient's basic condition and monitor the progress of the disease.Symptom information is also listed in the clinic information form, which helps doctors comprehensively understand the patient's condition, predict disease progression, and take relevant measures for intervention. Domain experts can view this information by clicking on the drop-down list.} 


\subsubsection{Information Visualization Module}
\label{sec:Visualization display module}

\textcolor{blue}{
To explore the statistical relationship between physiological indicators of patients and the plaque volume (represented by voxels) on the left and right sides, the system integrated a series of linked information visualization charts (\textbf{DG2}). Specifically, we display the relationships between four carotid artery plaque components (IPH, IPH\_lipid, Calcium, and Fibrous) with physiological indicators. Each chart is interactive, allowing users to hover over data points to directly display specific numerical values of the variables. When hovering over a specific component, other components are faded to focus on the distribution patterns and trends of that particular component. Users can also hide or show a component by clicking on the legends. Data points colors in the information visualization components represent carotid artery plaque components. Details of the information visualizations are described below.}


\vspace{3pt}
\noindent\textbf{1) Age vs Plaque Size.}
\begin{color}{blue}
    To explore the relationship between age and carotid artery plaque components, we use a stacked area chart. The vertical axis represents the size of the plaques, while the horizontal axis represents age groups. It allows vascular surgeons to gain an intuitive understanding of the development of carotid artery plaques along with ages. It also highlights the overall trends of the four plaque components and gives users a straightforward comparison between different plaque components. It can be seen in "AGE vs P.size" in Fig.~\ref{fig:main}B. 
\end{color}

\vspace{3pt}
\noindent\textbf{2) BMI vs Plaque Size.}
\begin{color}{blue}
    BMI are associated with indices of carotid stiffness and plaque volume among Type II diabetes mellitus~\cite{botvin2020long}.Therefore, the line chart explores the relationship between BMI and four carotid artery plaque components. The vertical axis represents the voxels of carotid artery plaque components, while the horizontal axis represents BMI, which ranges from 17 to 30. One could drag the mouse to the specific voxel numbers at the corresponding positions on this view, as seen in "BMI vs P.size" in Fig.~\ref{fig:main}B. 
\end{color}

\noindent\textbf{3) Gender vs Plaque Components.}
\begin{color}{blue}
    Pie charts are used to represent the relationship between gender and the proportion of plaque components. It specifically highlights the proportion of different components represented by different colors within each gender and compares the differences in plaque components between males and females, as seen in "GNDR vs P.cmp" in Fig.~\ref{fig:main}B.
\end{color}

\vspace{3pt}
\noindent\textbf{4) BNP \& TN vs Plaque Size.}
TN and BNP are two biomarkers commonly used for diagnosing and monitoring heart diseases, which has strong effects on atherosclerotic plaques.~\cite{omland2008circulating}. 
\begin{color}{blue}
    Therefore, it is necessary to explore the internal relationship between the two factors and the carotid artery plaque components. By using a scatter plot, in which the horizontal axis shows the range of BNP values and the vertical axis shows the size of plaque components. Different carotid artery plaque components are represented in different colored circles in the scatter plot. This enables users to gain a macro perception of overall distribution of BNP values and plaque sizes, which can be seen in "BNP vs P.size" in Fig.~\ref{fig:main}B. Similarly, using a scatter plot users can identify the TN distributions of carotid artery plaque components along with plaque sizes. 
\end{color}

\begin{color}{blue}
    \vspace{3pt}
    \noindent\textbf{5) Chronic Diseases vs Plaque Components.} The stacked bar charts of "CHR vs P.cmp" in Fig.~\ref{fig:main}B illustrates the relationship of chronic diseases and carotid artery plaque components, which specifically focus on plaque components voxels and the number of patients. 
    The upper part of the stacked bar chart represents the average size of carotid artery plaques for different chronic diseases. It provides users an overview understanding of the impacts of carotid artery plaque components to chronic diseases. The lower part of the stacked bar chart represents the number of patients with different plaque components under a specific chronic disease. It is important to note that a patient may have multiple plaque components. 
\end{color}

\subsubsection{Multiclass Classification Module}
\label{sec:plaque Category Prediction Module}
This module takes the data described in Sec.\ref{sec:dataDescription} as input and utilizes multiclass classification model, Support Vector Machine (SVM), to evaluate the importance of various CTA imaging features and analyze the correlation between different plaque characteristics (\textbf{DG3}). 
SVM is a powerful supervised learning algorithm widely used in disease pattern recognition and classification tasks. 
Due to its excellent performance in medical tasks and the strong capability to handle high-dimensional data and small sample sets. We have chosen SVM as the target classifier. \textcolor{blue}{ In addition, in order to avoid over-fitting problem of the data samples, we performed $k$-fold cross-validation. Based on
our goals, the literature, and experiment, by default, we use
$k=5$, which we have found to strike a good balance. In each cross-validation iteration, we execute feature ranking and obtain the saliency measures. After averaging all cross-validation round, we obtain the top 10 features, which would be drawn in a parallel coordinate view below.}

\vspace{3pt}

\begin{color}{blue}
\noindent\textbf{1) Top Feature Correlation.}
This module displays the correlation of the top ten significant plaque features after executing machine learning model. The horizontal axis of the parallel coordinate view shows plaque features and the vertical axis shows the range of each plaque values. 
By observing this parallel coordinate view, outliers or values significantly deviating from other features can be quickly detected, indicating potential anomalies. By clicking on each feature, it pops out a performance metric table on the right-bottom of the system. One can directly obtain the ANOVA test results between the selected feature and chronic diseases along with the corresponding P-value and V-value for each feature(\textbf{DG4}). Meanwhile, a bar chart shows ANOVA F-value of each feature and chronic disease would be plotted, which helps in determining whether there is a significant association between plaque components and each chronic disease, and assess the impact of chronic diseases on that particular feature. 

\vspace{3pt}
\noindent\textbf{2) Top Feature Distribution.}
To demonstrate the performance of different features, we used two views to visualize the top 10 features: radar map and boxplot. Users can view it by selecting the "Top Feature Distribution" drop-down box in Fig.\ref{fig:main}C. When drawing a radar map, we map features from multiple dimensions onto a coordinate axis, where each dimension's feature corresponds to a coordinate axis and is arranged radially at the same spacing. Meanwhile, multiple coordinate axes are unified into a single metric for normalization, this helps to show the weight of the top 10 features. 
The radar view also helps in viewing the performance of each feature on different carotid artery plaque components, which leads to a better understanding of the similarity and difference of carotid artery plaques. In addition, a boxplot view has been designed to show the distribution of top 10 feature values. 
One can intuitively obtain the distribution of different carotid plaque components and can effectively evaluate the feature deviation. This helps in infer the impact of features on specific carotid plaque components and promote further attention and investigation on outliers.

\vspace{3pt}
\noindent\textbf{3) ROC\& PR Curve.}
Receiver operating characteristic curve (ROC) curve and Precision-Recall (PR) Curve curve are used to measure the performance of classifiers in classification. ROC curve can be used to evaluate the effectiveness of carotid artery plaque components in disease analysis. By examining the ROC curves of different plaque components, one can assess their correlation with the target condition.
This is valuable in guiding clinical decisions and improving diagnostic accuracy. Similarly, PR curve can be used to evaluate the performance of different carotid artery plaque components in diagnosing pathological conditions. By calculating and plotting the precision and recall of the model at different classification thresholds. It provides a comprehensive assessment of the classification model. 
A component's curve is closer to the upper-right corner and has a higher AUC value indicates the corresponding component might be an effective diagnostic indicator.

\end{color}


\subsubsection{Carotid Artery Plaque Rendering Module}
\label{sec:plaque Display Module}

\begin{figure*}
    \centering
    \includegraphics[width=\textwidth]{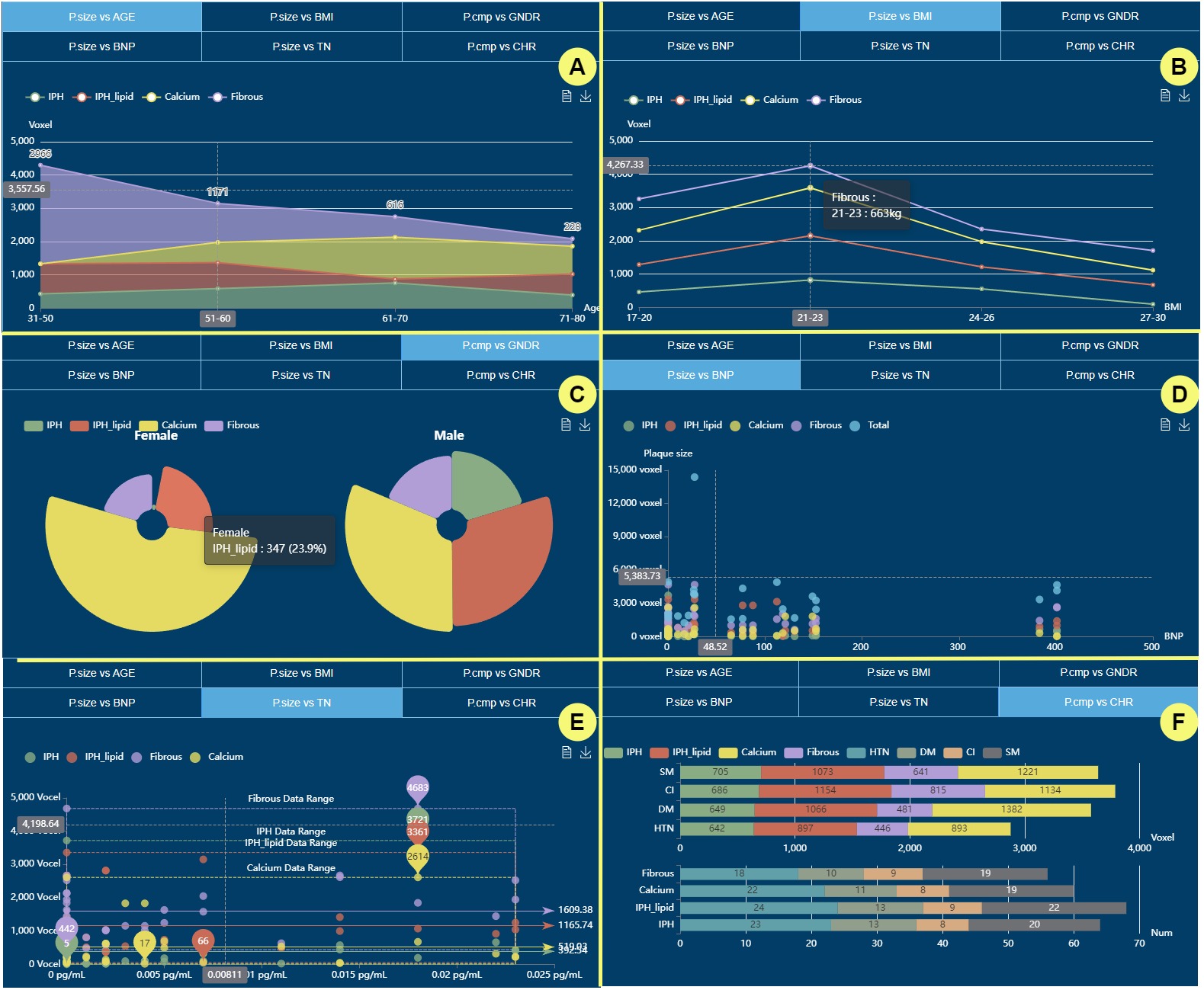}
    \caption{Performing Case Study 1 with the VA system.Here we show the Plaque size vs Age(A), Plaque Size vs BMI (B), Plaque Composition vs Gender (C), Plaque Size vs BNP (D), Plaque Size vs Troponin (E), Plaque Composition vs Chronic (F).}
    \label{fig:case1}
\end{figure*}

\begin{color}{blue}
Since vascular surgeons are accustomed to observing CTA images in practical work and they require a detail visualization of carotid artery plaque components. It is needed to implement the ability to display carotid artery CTA images and intuitively showcase plaque components in the system (\textbf{DG1, DG4}). Fig.~\ref{fig:main}D shows the rendering module of the proposed carotid artery plaque visualization system. The left side of Fig.~\ref{fig:main}D is CTA images interaction module, which contains a list of image processing tools, including image transformation, enhancement, and measurement.etc. It has a range of data interaction capabilities, such as scaling, translation, rotation, and flipping, which enables clinicians to freely adjust the view and position of the image. 
The right side of Fig.~\ref{fig:main}D shows the plaque components visualization images, in which different carotid artery plaque components are rendered in different colors. It has a visual magnification capability to assist vascular surgeons in observing the details of plaques more closely. 
\end{color}

\section{Evaluation}
\label{sec:cases}

\textcolor{blue}{
To evaluate our system, we have provided case studies and received feedback from domain experts. Two experts (E1, E2) with the vascular surgery background at Zhejiang Provincial People's Hospital were involved throughout the entire process, including system design, case studies, and qualitative feedback. E1 is a vascular specialist proficient in minimally invasive surgery and interventional treatment for various common vascular diseases. E2 is a chief physician of vascular surgery  with extensive clinical experience in various vascular diseases and inflammatory infections. The entire project lasted for about a year. We had meetings with domain experts every month, starting from conceptualization and system design. The experts provided feedback multiple times, leading us to revise the system until it met their basic requirements. Subsequently, they tested the functional modules of the system, provided a real dataset (22 subjects) 
for case analysis, and provide valable insights and suggestions based on the analysis results.
}


\subsection{Study 1: Exploration of Clinical Recording Data}
\begin{color}{blue}
The first requirement from the clinical experts is to analyze clinical record data, which includes conducting a comparative analysis between patients' demographics (age, gender, BMI) and four kinds of carotid plaque components or size, displaying the correlation between various chronic diseases (HTN, DM, Cl, SM) and the plaque components and size, as well as conducting correlation analysis between other relevant risk factors, e.g., TN and BNP, and the carotid plaque. Therefore, in our system, we have employed a series of information visualization techniques to present visualizations related to physiological indicators and plaque indicators, thereby it can intuitively represent group trends and the relationship between the individual and the group, as seen in Fig.~\ref{fig:case1}. 

\end{color}



\begin{color}{blue}
Firstly, we tried to identify the relationship between patients’ demographics and plaque indicators. As seen in Fig.~\ref{fig:case1}A, it can be easily to obtain the age distribution of patients. Fibrous plaque component is particularly significant and noticeably in the young age group (31-50) than any other age groups, which indicates that age might be one of the strong factors that have significant influence on plaque components. 
E1 said, age is a well-established risk factor for the development of carotid artery plaques, however, the exact mechanism by which age contributes to plaque development is not entirely understood. 
Beyond age, we also detect the correlations of BMI and plaque components, as seen in Fig.~\ref{fig:case1}B. The line chart clearly shows the trends of plaque components along with BMI values. 
The BMI values exist in between 21 and 23 might lead to larger plaque size. In addition, in the gender comparison pie chart (Fig.~\ref{fig:case1}C), one can clearly observed that both males and females have a larger proportion of Calcium and IPH\_lipid components in their plaques. 
In the scatter plot comparing BNP with plaque components(Fig.~\ref{fig:case1}D), we identified two patients with significantly higher BNP values compared to others, and one patient had an exceptionally large plaque area. We reported these outliers to the clinical physicians to draw attention to these special cases.

We then aimed at discovering the potential relation between chronic disease and plaques indicators, as well as the relevant correlations between other risk factor (TN and BNP) and plaque components. As seen in Fig.~\ref{fig:case1}D, the yellow circles (Calcium plaque component)generally lies on the bottowm of the coordinate system, which means Calcium plaque component have smaller size than other three types of plaque components. 
Similary, in Fig.~\ref{fig:case1}E, we can easily found it that most of the purple circles are on the top of orange circles. It indicates that the Fibrous plaques generally have larger size than the IPH plaques. Additionally, with the increase of TN value, there seems a slight increase in size. E2 said, actually TN itself doesn't directly influence the formation of carotid artery plaque, but elevated levels of TN in the bloodstream can indicate heart damage, which is often associated with atherosclerosis. In the bar chart of plaque components and chronic diseases(Fig.~\ref{fig:case1}F), we found that IPH\_lipid and Calcium components had higher proportions for different chronic diseases, and hypertension and smoking was more prevalent among patients with different plaque components.
\end{color}

\begin{figure*}
    \centering
    \includegraphics[width=\textwidth]{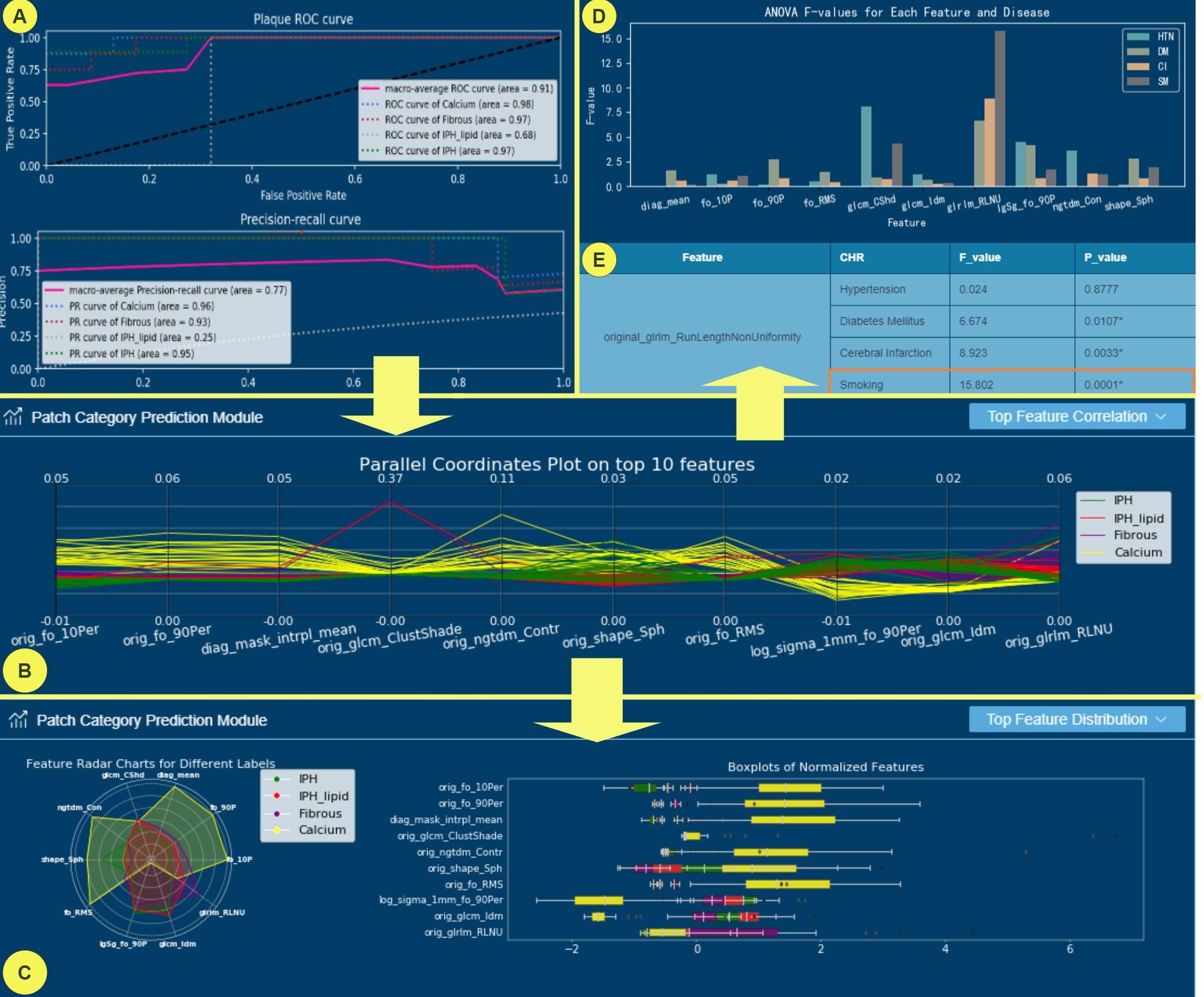}
    \caption{Performing Case Study 2 with the VA system.Here we show the machine learning performance(A), Top 10 feature correlation (B), Top 10 feature distribution (C), ANOVA analysis (D) and (E).}
    \label{fig:case2}
\end{figure*}

\subsection{Study 2: Integrated Exploration of Radiomic Features and Physiological Indicators}
\begin{color}{blue}
The second requirement proposed by vascular experts is to explore the carotid plaques in depth by combining radiomic features and physiological indicators to uncover inherent relationships. Therefore, we launched the second case study, as shown in Fig.~\ref{fig:case2}, which represents this analysis process.
  
Firstly, we conducted machine learning training on all the data. The performance of the multi-classification model training is shown in Fig.\ref{fig:case2}A, with high classification performance achieved for all types of plaque components. The top 10 important features from the model training are shown in Fig.\ref{fig:case2}B, where different colors represent different plaque components. From the figure, we can see that the Calcium component has a larger distinction from other plaque components in terms of the top 3 features. IPH$\_$lipid presents a different feature trend from other plaque components concerning the fourth feature. The detailed distribution of the top 10 importance features is shown in Fig.\ref{fig:case2}C. From the radar chart, it's not hard to see that the Calcium plaque component is drastically different from the other three in terms of feature distribution, attaining larger values in multiple features, while the values for IPH plaque component tend to be low. Simultaneously, from the boxplot, we can intuitively find that multiple feature values of Calcium far exceed other plaque components, but are significantly smaller than others for the \textit{'orig\_glrlm\_RLNU'} feature. E1 pointed out that it is necessary to further observe this plaque component and feature, which might be the most basic type of carotid artery plaque change. Then, Fig.\ref{fig:case2}D shows the ANOVA analysis results of the top 10 features. We can find that all carotid artery plaque components achieved a very high F-value for the \textit{'orig\_glrlm\_RLNU'} feature. Further, in the parallel coordinates plot, when we selected this feature, we found that its F-value related to smoking reached 15.802, and P-value was 0.0001. This proves that this feature is highly correlated with smoking. The impact of smoking on carotid artery plaques is likely to be reflected through this feature. E2 said that smoking is a significant risk factor for atherosclerosis, as smoking can damage the vessel walls and increase inflammation response, leading to atherosclerosis. 
  
\end{color}

\subsection{Expert Feedback}

\begin{color}{blue}
As stated, we sought feedback from cardiovascular Clinical specialists to evaluate our work. They provided valuable insights into the system's functionality, performance, usability, and offered insightful suggestions.

We implemented several functionalities based on the interactions with the domain experts. For example, we added the chronic disease relation views (e.g., Fig.~\ref{fig:case1}D, Fig.~\ref{fig:case1}E, and Fig.~\ref{fig:case1}F) based on E1’s request to investigate
how each chronic disease affect the carotid artery plaques. E1 affirmed the system's ability to intuitively display the correlation between various factors and carotid artery plaque. He believes that this visualization method streamlines intricate information by utilizing a variety of visual elements such as color, shape, and size. This approach not only makes the information more accessible and easier to comprehend but also fosters a deeper understanding of the underlying relationships. 
However, he also pointed out that while these intuitive visualization results could provide him with new knowledge and perspectives, broadening his horizon and stimulating his thinking, they are highly valuable in academic research and scientific exploration. In addition, E1 also explicitly stated that this does not mean they can be directly used for clinical diagnosis. Clinical diagnoses require validation of these correlations through large-scale clinical trials and research to confirm their existence and to understand how much impact they have on disease diagnosis and treatment,rather than relying solely on these visualization results. 

E2 is more focused on the correlation between plaque features and chronic diseases, believing that such correlations could provide important clues for understanding and preventing chronic diseases. He also explicitly mentioned that although machine learning has become an important tool in modern data analysis, many cardiovascular experts are not familiar with this technology and might be confused by complex models and parameter selection. Therefore, we hid the technical details of machine learning models and parameter selection.
For the ability to quickly find a high correlation between \textit{'orig\_glrlm\_RLNU'} and smoking in case 2 through interaction, E2 believes this ability is very helpful for identifying and understanding disease risk factors and guiding us to make more targeted and effective interventions. 
In addition, E2 offered constructive suggestions for our research. He believes that besides radiomic features, we could consider adding other measurement data for analysis, such as lipid levels, white blood cell count, lymphocyte count, and homocysteine. These biochemical indicators might have a significant impact on the correlation between plaque features and chronic diseases. He believes that integrating these data could yield very valuable disease knowledge, providing a more comprehensive and accurate reference for diagnosis and treatment.Their professional knowledge and comments are insightful. We consider their suggestions will assist VA researchers in better planning and executing future work.

\end{color}


\section{Discussion and Limitations}

\textcolor{blue}{
Our system effectively aids vascular surgeons in the efficient diagnosis of carotid artery plaques by integrating patient physiological indicators and radiomic features. Using a series of visualization components, domain experts can observe disease patterns among populations and explore pathophysiological mechanisms of carotid atherosclerosis.}

\textcolor{blue}{
The scarcity of data is the principal constraint in our work, which notably narrows the range of visual chart types we can include. Currently, we are confined to comparing plaque components as a characteristic feature of plaque with only a handful of physiological indicators, leading to a restricted scope in disease analysis. This paucity of data also inhibits our ability to explore underlying patterns within the indicators. Moving forward, we will maintain collaboration with domain experts in the hospital to ensure that the data in the system is updated in a timely manner. Another limitation is that we lack temporal data, preventing us from observing the comparative changes before and after treatment by analyzing the differences in the changes of carotid artery plaques over time. We hope to conduct follow-up studies on patients, and validate our analysis results through extensive temporal data. Another constraint involves the color overlap in some visualization views. While experts have pointed out this concern, their emphasis is more on the need for enhanced interactive analysis between radiomic features and physiological indicators. Recognizing both aspects, we plan to explore entirely new approaches in information visualization in our future research to address these interconnected challenges.}

\section{Conclusions}
\textcolor{blue}{Our system integrates patients' physiological indicators and radiological features, utilizing visual views and machine learning classification models to assist domain experts in better observing the distribution patterns of disease data. This helps in further exploring the pathophysiological mechanisms of carotid atherosclerosis. The system provides doctors with more clinical insights, enhances their confidence in disease identification, and effectively aids vascular surgeons in the accurate diagnosis of carotid artery plaques.}


%





\ifCLASSOPTIONcaptionsoff
  \newpage
\fi



\bibliographystyle{IEEEtran}
\bibliography{IEEEabrv,00_bibliography}

%



%







\vfill


\end{document}